\documentclass[a4paper,11pt]{iopart}
\usepackage{graphicx}
\usepackage{cite}
\usepackage{color}
\usepackage{amsfonts}
\usepackage{srcltx}
\begin{document}

\title{Entropy production and fluctuation relations for a KPZ interface}
\author{A C Barato$^1$, R Chetrite$^2$, H Hinrichsen$^1$ and D Mukamel$^2$}
\address{$^1$Universit\"at W\"urzburg,
     Fakult\"at f\"ur Physik und Astronomie\\
         D-97074 W\"urzburg, Germany\\
    $^2$Department of Physics of Complex Systems,
    Weizmann Institute of Science\\
    Rehovot 76100, Israel}

\ead{barato@physik.uni-wuerzburg.de}

\def\ex#1{\langle #1 \rangle}

\begin{abstract}
We study entropy production and fluctuation relations in the
restricted solid-on-solid growth model, which is a microscopic
realization of the KPZ equation. Solving the  one dimensional
model exactly on a particular line of the phase diagram we
demonstrate that entropy production quantifies the distance from
equilibrium. Moreover, as an example of a physically relevant
current different from the entropy, we study the symmetry of the
large deviation function associated with the interface height. In a
special case of a system of length $L=4$ we find that the
probability distribution of the variation of height has a symmetric
large deviation function, displaying a symmetry different from the
Gallavotti-Cohen symmetry.
\end{abstract}

\section{Introduction}

The theoretical understanding of out of equilibrium systems is one
of the main goals in statistical physics. Nonequilibrium phenomena
display very rich behavior, with several features not observed in
equilibrium, as for example phase transitions in one-dimensional
systems with short-range interactions. Recently great theoretical
progress has been made concerning systems in contact with two
reservoirs \cite{bertini01,bertini02,bodineau04,derrida07,tailleur08}.
Nevertheless, for general nonequilibrium systems very little is
known.

For some continuous time Markov jump processes an (average) entropy
production can be defined \cite{schnakenberg76,seifert05}. This
quantity is zero if detailed balance is fulfilled, which corresponds
to an equilibrium stationary sate, and it is positive in the case of
a nonequilibrium stationary state. Therefore, a positive entropy
production in the stationary state is a signature of nonequilibrium.
In spite of the generality of this statement the relation between
entropy production and nonequilibrium stationary states is still
poorly understood \cite{zia07}. Recent studies in this direction
concern the relation between entropy production and nonequilibrium
phase transitions, where the entropy production as a function of the
control parameter has been observed to peak near the critical point
\cite{gaspard04,tome,andrae10}.

In this paper we will consider a microscopic surface growth model in
the Kardar-Parisi-Zhang \cite{KPZ} (KPZ) universality class. The KPZ
equation reads
\begin{equation}
\partial_t h(x,t)= v+ \nu\nabla^2h(x,t)+ (\lambda/2)(\nabla h(x,t))^2+\eta(x,t),
\end{equation}
where $x$ denotes the position on a $d-$dimensional interface (we
will be dealing only with the $d=1$ case), $h(x,t)$ represents the
height of the interface and $\eta(x,t)$ is a Gaussian white noise.
The Laplacian term is related to surface tension, $v$ is the
velocity of the interface at zero slope and the non-linear term is
the lowest order term that breaks the up-down symmetry
($h(x,t)\to-h(x,t)$) \cite{barabasi95,krug97}. Assuming that a
nonequilibrium stationary state of a KPZ interface is characterized
by $v$ and $\lambda$, we study how the entropy production varies
with these parameters. We shall argue that this is a reasonable
assumption and it allows us to investigate how entropy production
characterizes a nonequilibrium stationary state. We will work with
the restricted solid on solid (RSOS) growth model, because this
model has a rich phase diagram and we can perform analytical
calculations (exact and approximative).

The positivity of the average entropy production is a statement
analogous to the second law of thermodynamics. More generally, it is
possible to associate a fluctuating entropy with a single stochastic
path \cite{seifert05}, and the probability distribution of this
fluctuating entropy is constrained through the fluctuation relation
\cite{evans93,evans94,gallavotti95,kurchan98,lebowitz99,maes99,
Jia1,andrieux07,harris07,kurchan07}, which is a stronger statement
than the positivity of the average entropy production. The
fluctuation relation implies the existence of a symmetry of the
large deviation function associated with the probability
distribution of entropy which is known as the Gallavotti-Cohen (GC)
symmetry \cite{gallavotti95,lebowitz99,harris07}.

In general, unlike the entropy production, other time-integrated
currents are not expected to have symmetric large deviation function
\cite{lebowitz99}.  In a growing interface, the physically relevant
time-integrated current is the variation of the interface height.
The main focus of this work is to analyze the
symmetry of the large deviation function associated with the
variation of height using the RSOS model.  We find that while
indeed no symmetry of the large deviation function exists in the
general case, the large deviation function related to the variation
of height is symmetric in the case of a four-site system. This is in
spite of the fact that entropy production and height are not
proportional to each other in the long time limit. Unlike the GC symmetry which results from a certain relation between the rate of each
microscopic trajectory and its time reversed one, this symmetry is a result of
a similar relation between the rates associated with certain groups of trajectories. 

The paper is organized as follows. In Sec. 2 we define entropy and briefly discuss its relation with nonequilibrium stationary states. In Sec. 3 we define the RSOS model and calculate the average entropy production analytically, analyzing how it depends on the interface velocity and $\lambda$. The fluctuation relation and the GC symmetry are introduced in Sec. 4. In Sec. 5 we investigate fluctuations of the interface height. We show that if a second time-integrated current is considered a fluctuation relation for the joint probability distribution of two currents, with one of the currents being the height, can be found and we also show that for the four-site system the large deviation function related to height is symmetric, displaying a symmetry different from GC symmetry. We conclude in Sec. 6. 

\section{Entropy production in nonequilibrium stationary states}

In the following discussion we consider continuous time Markov
processes that obey the master equation
\begin{equation}
\frac{d}{dt}P(C,t)= \sum_{C\neq C'}[w_{C'\to C}P(C',t)- w_{C\to C'}P(C,t)],
\label{c5eqmaster1}
\end{equation}
where $P(C,t)$ is the probability of being in state $C$ at time $t$
and $w_{C\to C'}$ is the transition rate from state $C$ to $C'$. For
such stochastic processes three different fluctuating entropies can
be defined, namely \cite{seifert05}
\begin{quote}
\begin{itemize}
\item[(a)] the \textit{configurational entropy} of the system $S_{sys}(t)=-\ln P(C,t)$,
\item[(b)] the \textit{entropy of the medium} (=environment) $S_m(t)$ which changes\\ instantaneously by $\ln(w_{C\to C'}/w_{C'\to C})$ during a transition from $C$ to $C'$.
\item[(c)] and the \textit{total entropy} $S_{tot}(t)=S_{sys}(t)+S_m(t)$.
\end{itemize}
\end{quote}
Note that the definition of the environmental entropy requires the
rates to satisfy the condition that if $w_{C\to C'}>0$ then the
reverse rate $w_{C'\to C}$ has to be positive as well.

A stochastic path is a sequence of transitions
\begin{equation*}
C_0 \to C_1 \to C_2 \to \ldots C_N\,,
\end{equation*}
taking place at random instances of time $T_0 \le t_1<t_2<\ldots<t_N \le T$
according to the specific rates during a fixe time interval $\Delta
T= T-T_0$. Such a path of
transitions changes the aforementioned entropies by
\begin{eqnarray}
\Delta{S}_{sys} &=& \ln P(C_0,t_0)- \ln P(C_N,t_N)\,,
\label{c5defentropy2}\\
\Delta{S}_{m} &=& \sum_{i=1}^N \ln\frac{w_{C_{i-1}\to C_i}}{w_{C_{i}\to C_{i-1}}}\,,
\label{c5defentropy}
\end{eqnarray}
and $\Delta{S}_{tot} = \Delta{S}_{sys} + \Delta{S}_{m}$. While these
entropies vary discontinuously for a particular stochastic path,
their \textit{expectation values} vary smoothly according to
\begin{eqnarray}
\dot s_{sys}(t)\;:=\;\frac{d}{dt}\ex{S_{sys}(t)} &=& \sum_{C,C'}\ln\frac{P(C,t)}{%
P(C',t)}P(C,t)w_{C\to C'}\,, \label{c5sprods}\\
\dot s_m(t) \;:=\; \frac{d}{dt}\ex{S_{m}(t)} &=& \sum_{C,C'}\ln\frac{w_{C\to C'}}{w_{C'\to C}}P(C,t)w_{C\to C'}\,,
\label{c5sprod}\\
\dot s_{tot}(t) \;:=\; \frac{d}{dt}\ex{S_{m}(t)} &=& \sum_{C,C'}\ln\frac{P(C,t)w_{C\to C'}}{P(C',t)w_{C'\to C}}P(C,t)w_{C\to C'}\,,
\end{eqnarray}
where $\ex\ldots$ denotes the ensemble average. Since the last
equation can be rewritten as
\begin{equation}
\label{stot}
\fl \dot s_{tot}(t) = \frac12\sum_{C,C'}
\Bigl( \ln[P(C,t)w_{C\to C'}] - \ln[P(C',t)w_{C'\to C}]\Bigr)\, \Bigl(P(C,t)w_{C\to C'}-P(C',t)w_{C'\to C}\Bigr)
\end{equation}
and both brackets have always the same sign, the average total
entropy can only increase and will stay constant if detailed balance
is satisfied, in accordance with the second law of thermodynamics.
Moreover, if the system is in a non-equilibrium stationary state,
the average internal entropy $\ex{S_{sys}(t)}=-\sum_CP(C)\ln P(C)$
is constant so that $\dot s_{tot}=\dot s_m$. Since the stationary
master equation
\begin{equation}
0 \;=\; \sum_{C'}w_{C'\to C}P(C')-\sum_{C'}w_{C'\to C}P(C)
\end{equation}
is invariant under the replacement $w_{C\to C'} \to w_{C\to
C'}+A(C,C')/P(C)$ with an arbitrary symmetric function
$A(C,C')=A(C',C)$ while Eq.~(\ref{stot}) is not invariant under this
operation, the same stationary state may be generated by different
dynamical rules with a different entropy production rate (see
\cite{zia07}).

\section{Entropy production in a solid-on-solid growth model}

\subsection{Definition of the model}

\begin{figure}[t]
\centering\includegraphics[width=90mm]{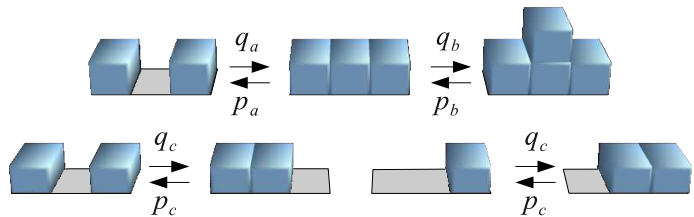}
\vspace{-3mm}
\caption{Deposition and evaporation rates for the RSOS model.}
\label{fig:rules}
\end{figure}

As a microscopic realization of the KPZ equation let us consider a
restricted solid-on-solid (RSOS) growth model on a one-dimensional
lattice with $L$ sites and periodic boundary conditions. The
configuration $C$ of the interface is characterized by height
variables $h_i\in\mathbb{Z}$ attached to the lattice sites $i$ which
obey the restriction
\begin{equation}
\label{rsos}
|h_i-h_{i\pm 1}| \leq 1\,.
\end{equation}
The model evolves random-sequentially by deposition and evaporation
of particles with rates $q_{a,b,c}$ and $p_{a,b,c}$ according to the
rules shown in Fig.~\ref{fig:rules}. Because of (\ref{rsos}), it is
convenient to describe the configurations of the interface in terms
of charges
\begin{equation}
\label{charges}
\sigma_i\;:=\; h_{i+1}-h_i \,\in \, \{0,\pm 1\}
\end{equation}
which evolve random-sequentially according to the rules

    \begin{equation}
    \label{chargerules}
    \end{equation}
    \vspace{-11mm}
    \setlength{\unitlength}{4144sp}%
    \begingroup\makeatletter\ifx\SetFigFont\undefined%
    \gdef\SetFigFont#1#2#3#4#5{%
      \reset@font\fontsize{#1}{#2pt}%
      \fontfamily{#3}\fontseries{#4}\fontshape{#5}%
      \selectfont}%
    \fi\endgroup%
    \begin{center}\begin{picture}(3412,433)(20,256)
    \thinlines
    {\color[rgb]{0,0,0}\put(220,520){\vector( 1, 0){382}}
    }%
    {\color[rgb]{0,0,0}\put(602,451){\vector(-1, 0){382}}
    }%
    {\color[rgb]{0,0,0}\put(880,520){\vector( 1, 0){382}}
    }%
    {\color[rgb]{0,0,0}\put(1262,451){\vector(-1, 0){382}}
    }%
    {\color[rgb]{0,0,0}\put(1957,520){\vector( 1, 0){383}}
    }%
    {\color[rgb]{0,0,0}\put(2340,451){\vector(-1, 0){383}}
    }%
    {\color[rgb]{0,0,0}\put(3000,520){\vector( 1, 0){382}}
    }%
    {\color[rgb]{0,0,0}\put(3382,451){\vector(-1, 0){382}}
    }%
    \put(672,451){\makebox(0,0)[lb]{\smash{{\SetFigFont{9}{10.8}{\rmdefault}{\mddefault}{\updefault}{\color[rgb]{0,0,0}$00$}%
    }}}}
    \put(1312,451){\makebox(0,0)[lb]{\smash{{\SetFigFont{9}{10.8}{\rmdefault}{\mddefault}{\updefault}{\color[rgb]{0,0,0}$+-$}%
    }}}}
    \put(324,312){\makebox(0,0)[lb]{\smash{{\SetFigFont{9}{10.8}{\rmdefault}{\mddefault}{\updefault}{\color[rgb]{0,0,0}$p_a$}%
    }}}}
    \put(324,590){\makebox(0,0)[lb]{\smash{{\SetFigFont{9}{10.8}{\rmdefault}{\mddefault}{\updefault}{\color[rgb]{0,0,0}$q_a$}%
    }}}}
    \put(984,312){\makebox(0,0)[lb]{\smash{{\SetFigFont{9}{10.8}{\rmdefault}{\mddefault}{\updefault}{\color[rgb]{0,0,0}$p_b$}%
    }}}}
    \put(984,590){\makebox(0,0)[lb]{\smash{{\SetFigFont{9}{10.8}{\rmdefault}{\mddefault}{\updefault}{\color[rgb]{0,0,0}$q_b$}%
    }}}}
    \put(2062,312){\makebox(0,0)[lb]{\smash{{\SetFigFont{9}{10.8}{\rmdefault}{\mddefault}{\updefault}{\color[rgb]{0,0,0}$p_c$}%
    }}}}
    \put(2062,590){\makebox(0,0)[lb]{\smash{{\SetFigFont{9}{10.8}{\rmdefault}{\mddefault}{\updefault}{\color[rgb]{0,0,0}$q_c$}%
    }}}}
    \put(3104,312){\makebox(0,0)[lb]{\smash{{\SetFigFont{9}{10.8}{\rmdefault}{\mddefault}{\updefault}{\color[rgb]{0,0,0}$p_c$}%
    }}}}
    \put(3104,590){\makebox(0,0)[lb]{\smash{{\SetFigFont{9}{10.8}{\rmdefault}{\mddefault}{\updefault}{\color[rgb]{0,0,0}$q_c$}%
    }}}}
    \put(2811,451){\makebox(0,0)[lb]{\smash{{\SetFigFont{9}{10.8}{\rmdefault}{\mddefault}{\updefault}{\color[rgb]{0,0,0}$-0$}%
    }}}}
    \put(1744,441){\makebox(0,0)[lb]{\smash{{\SetFigFont{9}{10.8}{\rmdefault}{\mddefault}{\updefault}{\color[rgb]{0,0,0}$0+$}%
    }}}}
    \put(2382,451){\makebox(0,0)[lb]{\smash{{\SetFigFont{9}{10.8}{\rmdefault}{\mddefault}{\updefault}{\color[rgb]{0,0,0}$+0$}%
    }}}}
    \put(3417,451){\makebox(0,0)[lb]{\smash{{\SetFigFont{9}{10.8}{\rmdefault}{\mddefault}{\updefault}{\color[rgb]{0,0,0}$0-$}%
    }}}}
    \put( 0,451){\makebox(0,0)[lb]{\smash{{\SetFigFont{9}{10.8}{\rmdefault}{\mddefault}{\updefault}{\color[rgb]{0,0,0}$-+$ }%
    }}}}
    \end{picture}\end{center}%
    \vspace{-2mm}

\noindent Since deposition and evaporation correspond to negative
and positive charge displacements, the average interface velocity
$v=\frac d{dt} \ex{h_i}$ is given by
\begin{equation}
\label{eq:velocity}
\fl v = (q_b-p_a) \ex{00} \,+\, q_0 \ex{-+}  \,-\, p_b \ex{+-} \,+\, q_c \bigl( \ex{0+}+\ex{-0} \bigr) \,-\, p_c \bigl( \ex{+0}+\ex{0-} \bigr)\,,
\end{equation}
where $\ex{\sigma_i\sigma_{i+1}}$ is the probability to find two
specific charges at neighboring sites. Likewise the entropy exported
to the environment is given by
\begin{eqnarray}
\label{eq:entropypair}
\dot s_m \;&=&
\Bigl(q_b\ln\frac{q_b}{p_b}-p_a\ln\frac{q_a}{p_a} \Bigr) \ex{00} \;+\;
q_a\ln\frac{q_a}{p_a} \ex{-+} \;-\; p_b\ln\frac{q_b}{p_b}\ex{+-} \nonumber \\
&& + \; q_c \ln\frac{q_c}{p_c}\Bigl(\ex{0+}+\ex{-0}\Bigr) \;-\;
p_c\ln\frac{q_c}{p_c}\Bigl( \ex{+0}+\ex{0-} \Bigr)\,.
\end{eqnarray}

The interface slope $m$ is defined by
\begin{equation}
m= \frac1L\sum_{i=1}^L\sigma_i.
\end{equation}
Since we consider periodic boundary conditions we have $m=0$ and,
therefore, the number of positive charges is equal to the number of
negative charges. One way of introducing a non-zero interface slope
in the RSOS model is to change the RSOS restriction between sites
$L$ and $1$ so that $h_L-h_1= H, H\pm1$. In this case the interface
slope is given by $m= H/L$. It can be shown that the $\lambda$
term in the KPZ equation can be defined for a microscopic model. It
is related to how the velocity depends on the interface slope in the
following way \cite{barabasi95},
\begin{equation}
v(m)= v+ (\lambda/2)m^2.  \label{c5eqlambda2}
\end{equation}

We assume that a nonequilibrium stationary state of the RSOS model is characterized by $v$ and
$\lambda$, this means that it is characterized by the interface velocity and
by how the velocity depends on the interface slope. In the following we
calculate the entropy production, the velocity and the parameter $\lambda$
defined above. We are interested in investigating how the entropy production
is a function of these two quantities.

\subsection{Entropy production: Exact results}

\begin{figure}[t]
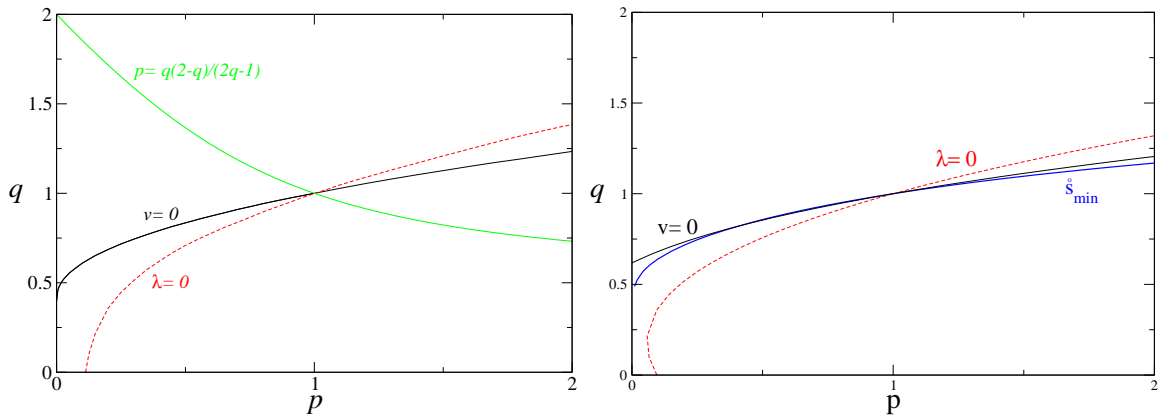

\centering
\includegraphics[width=75mm]{Fig2a.eps}
\includegraphics[width=75mm]{Fig2b.eps}
\vspace{-3mm} \caption{Left: Phase diagram of
the RSOS model obtained from numerical simulations and the line (\ref{eq:exactline}) where the model is exactly
solvable. Right: Mean field phase diagram (see text).} 
 \label{c5figmf}
\end{figure}

As shown in Appendix A, the RSOS model can be solved exactly in the
thermodynamic limit $L \to \infty$ in a certain parameter subspace
\cite{neergaard97}. To simplify the analysis, we will from now on
restrict ourselves to a two-dimensional section of the parameter
space \cite{hinrichsen97}
\begin{equation}
q_a=q_b=q_c=q\qquad p_a=p\qquad p_b=p_c=1
\end{equation}
controlled by two parameters $q$ and $p$. In this section the
exactly solvable subspace corresponds to the line
\begin{equation}
\label{eq:exactline}
p=\frac{2q-q^2}{2q-1}\,.
\end{equation}
As shown in the phase diagram in the left panel of
Fig.~\ref{c5figmf}, this line intersects with the lines for $v=0$
and $\lambda=0$ at the point $p=q=1$, where the model can be shown
to obey detailed balance~\cite{hinrichsen97}. Note that the lines
$v=0$ and $\lambda=0$ intersecting at the detailed balance point is
in agreement with the assumption that $v$ and $\lambda$ characterize
the nonequilibrium stationary state of the RSOS model.

Along the exactly solvable line, the density of charges (cf.
Eq.~(\ref{c5eqrho})) is given by
\begin{equation}
\rho= \frac{2}{2+\sqrt{(2q-1)/q}}.
\end{equation}
As shown in appendix B, the two-point function in the charge basis factorizes in the exactly solvable line, so that the velocity $v$, the nonlinearity coefficient $\lambda$,
and the entropy production rate $\dot s_m$ are given by
\begin{eqnarray}
v &=& (q-1)\rho\,,  \label{c5vforplot}\\
\lambda &=& -(q - 1)\frac{\rho}{4q(1-\rho)}\Bigl[1+6q-2\rho(2q+1)\Bigr],
\label{c5lamforplot}
\\
\dot{s}_m&=& (q-1)\Bigl(\frac{\rho^2}{2}\ln\frac{2q-1}{2-q}+\frac{\rho}{2}%
(2-\rho)\ln q\Bigr).  \label{c5sforplot}
\end{eqnarray}

\begin{figure}[t]
\centering
\includegraphics[width=75mm]{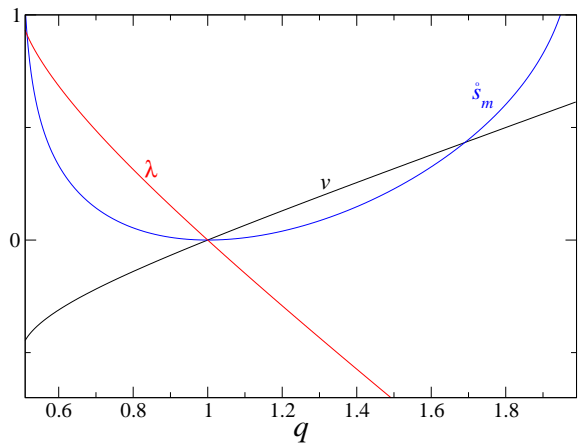}
\vspace{-3mm}
\caption{$v(q),\lambda(q)$ and $\dot s_m(q)$ along the exactly solvable line.}
\label{c5figplot}
\end{figure}

In Fig \ref{c5figplot} we plot the three quantities as a function of $q$. We
see that the entropy production is always positive and it grows as the
distance from the equilibrium point $q=1$ increases. The velocity and $%
\lambda$ have opposite signs and as their absolute values increase
the entropy production increases. This is in agreement with the idea
that the entropy production gives a measure of how far the system is
from equilibrium.

\subsection{Entropy production: Approximate results}

The two-point function in the charge basis
factorizes only on the line $p=q(2-q)/(2q-1)$. In order to access
the whole phase diagram we now assume that it factorizes for any
$p,q$, which is analogous to a pair-mean field approach in the
height representation \cite{barato07}. Details of the
calculation are presented in Appendix B. Within this mean-field
approximation, from (\ref{eqB5}), the density of charges in the
stationary state (at zero slope) is given by
\begin{equation}
\rho= 2/(2+{((q+p)/(q+1))}^{-1/2})\,.  \label{c5eqrhomf}
\end{equation}
Repeating the above calculations we find that the interface velocity
vanishes for
\begin{equation}
p= -q+q^2+q^3.  \label{c5v=0}
\end{equation}
and that the coefficient of the non-linear term in the KPZ equation vanishes for
\begin{equation}
\fl p= (-3 - q + 5 q^2 + q^3\pm\sqrt{5 + 10 q - 9 q^2 - 20 q^3 + 3 q^4 + 10 q^5
+ q^6})/8\,,  \label{c5lam=0}
\end{equation}
where the sign in front of the square root is positive for $p>1$ and
negative for $p<1$. As shown in the right panel of Fig.
\ref{c5figmf}, these lines approximate the true lines, cross at the
equilibrium point, and have the same type of curvature. In addition,
the figure shows the line where the entropy production obtained form equations (\ref{c5eqrhomf}) and (\ref{c5eqentprod}), is minimal. This line does not coincide with the line
$v=0$ although it is very close to it. Similar situations have been
observed in other models \cite{gaspard04,tome,andrae10}, where the
entropy production was found to peak near the critical point. In the
present case, this happens because although the entropy production
strongly depends on $v$, it also depends on $\lambda$.

In principle $\dot{s}_m$ can be considered functions of
$v$ and $\lambda$. However, we could find analytical expressions
only for $v=0$, where $p= -q+q^2+q^3$. On this line, from
(\ref{c5eqrhomf}) and (\ref{c5eqlambda})  we find
\begin{equation}
\lambda= (q-1)/q.
\end{equation}
For large $q$ we have $\lambda\to 1$ and at $q=(\sqrt5-1)/2)$, which
corresponds to $p=0$, $\lambda= -(\sqrt5-1)/2\approx
-0.618$. From (\ref{c5eqentprod}) we have
\begin{equation}
\dot{s}_m= -\frac{\lambda}{(\lambda-3)^2}\ln\bigg(\frac{(\lambda-1)^3}{\lambda^2-\lambda-1}\bigg).
\end{equation}
where the above relation is valid for $-(\sqrt5-1)/2<\lambda<1$. As
expected, the entropy production is zero at $\lambda=0$ and grows as
the absolute value of $\lambda$ grows. Note that it diverges for
$q\to (\sqrt5-1)/2)$ and $q\to \infty$. Again the results are in
agreement with the idea that entropy production gives a quantitative
measure of the distance from equilibrium.

\section{The Gallavotti-Cohen symmetry}
\label{flucsec}

The fluctuation relation states that
\begin{equation}
\frac{P(\Delta S_m)}{P(-\Delta S_m)}= \exp(\Delta S_m),
\label{c5infiniteFR}
\end{equation}
where $P(\Delta S_m)$ denotes probability of the variation of the
medium entropy during time $\Delta T$ and the above relation is
valid for $\Delta T\to \infty$. As we show below it leads to the GC
symmetry, which is a symmetry in the large deviation deviation
function associated with $P(\Delta S_m)$. In order to do that in a
more general framework we first introduce time-integrated currents
and the large deviation function associated with them.

Time-integrated currents are a functional of the stochastic path,
which changes only when the underlying process jumps to a different
configuration. They are defined by an increment $\theta_{C\to C'}$
that depends on the transition rates and is antisymmetric, i.e.
$\theta_{C\to C'}=-\theta_{C'\to C}$. As before we consider the path
$C_{0}\to C_{1}\to ...\to C_{N}$ starting at time $T_0$ and
finishing at time $T$, where $N$ is the total number of jumps and
$\Delta T= T-T_0$. A time-integrated current takes the generic form
\begin{equation}
J=\sum_{k=1}^{N}\theta _{C_{k-1}\rightarrow C_{k}}  \label{c5defJ}.
\end{equation}
In the case $\theta _{C\rightarrow C^{\prime }}=\ln
\frac{w_{C\rightarrow C^{\prime }}}{w_{C^{\prime }\rightarrow C}}$,
$J$ is just the entropy change $\Delta S_{m}$ (\ref{c5defentropy}).
Another current we are interested in is the variation of height,
where $\theta _{C\rightarrow C^{\prime }}$ is $+1$ for a deposition
and $-1$ for an evaporation.

The average of the time integrated current (\ref{c5defJ}) is
\begin{equation}
\left\langle J\right\rangle =\int_{0}^{\Delta T}dt
\sum_{C,C^{\prime }}\theta_{C\rightarrow C^{\prime }}P(C,t)w _{C\rightarrow C^{\prime }}
\end{equation}
In the long time limit, we have by ergodicity
\begin{equation}
\frac{ J }{\Delta T}\rightarrow \sum_{C,C^{\prime }}\theta _{C\rightarrow
C^{\prime }}P(C)w_{C\rightarrow C^{\prime }},
\end{equation}
where $P(C)$ is the stationary state probability distribution. The
large deviation function associated with this time-integrated
current is related to deviations of the current from its average
value. The large deviation function $I(j)$ is defined by
\cite{ellis85,touchette09,hollander}
\begin{equation}
 P(j) \approx \exp [-\Delta TI(j)],  \label{rate}
\end{equation}
where  $P(j)$ denotes the probability that $j= \frac{J}{\Delta T}$
and the above relation is valid for $\Delta T\to \infty$. The GC
symmetry is a symmetry of the large deviation function that reads
\cite{gallavotti95}
\begin{equation}
I(j)-I(-j)=-Ej
\label{ggc}
\end{equation}
The fluctuation relation for the external medium entropy
(\ref{c5infiniteFR}) corresponds to $E=1$ and was first proved, for
jump processes, by Lebowitz and Spohn \cite{lebowitz99}. The central
idea is to introduce the deformed time evolution operator
\begin{equation}
\widetilde{H}_{CC^{\prime }}(s)=\left\{
\begin{array}{c}
-w_{C^{\prime }\rightarrow C}\exp (-s\theta _{C^{\prime }\rightarrow C})%
\textrm{ if }C\neq C^{\prime } \\
\sum_{C^{\prime \prime }}w_{C\rightarrow C^{\prime \prime }}\textrm{ \ \ \ if }C=C^{\prime }
\end{array}
\right. .  \label{defmar}
\end{equation}
Note that for $s=0$ we get the usual time evolution operator for the
master equation. Consider the joint probability $P_J(C,T)$ of being
at a state $C$ at time $T$ with time-integrated current $J$. The
Laplace transform of it is
\begin{equation}
\tilde{P}_s(C,T)= \sum_J P_J(C,T)\exp(-sJ).
\end{equation}
From the master equation (\ref{c5eqmaster1}) we obtain \cite{lebowitz99}
\begin{equation}
\frac{d}{dT}\tilde{P}_s(C,T)= -\sum_{C'} \tilde{H}_{CC'}(s)\tilde P_s(C',T).
\end{equation}
Since $\tilde{H}(s)$ is a Perron-Frobenius operator, it is possible
to relate the generating function of the time-integrated current $J$
with its minimum eigenvalue in the following way,
\begin{equation}
\left\langle \exp (-sJ)\right\rangle \sim \exp (-\Delta T\hat{I}(s)),
\end{equation}
where $\hat{I}(s)$ is the minimum eigenvalue of the matrix
$\tilde{H}(s)$. The G\"artner-Ellis theorem
\cite{ellis85,touchette09,hollander} states that the large deviation
function (\ref{rate}) is the Legendre transform of $\hat{I}(s)$,
i.e.,
\begin{equation}
I(j)=\min_{s}\left( \hat{I}(s)-sj\right).
\end{equation}
Therefore the GC symmetry (\ref{ggc}) is equivalent to
\begin{equation}
\hat{I}(s)=\hat{I}(E-s).
\label{symvp}
\end{equation}
In the case where the time integrated current is the external medium
entropy (\ref{c5defentropy}), we have
\begin{equation}
\widetilde{H}_{CC^{\prime }}(s)=\left\{
\begin{array}{c}
-w_{C^{\prime }\rightarrow C}^{1-s}w_{C\rightarrow C^{\prime }}^{s}\textrm{ if
}C\neq C^{\prime } \\
\sum_{C^{\prime \prime }}w_{C\rightarrow C^{\prime \prime }}\textrm{ \ \ \ if }%
C=C^{\prime }
\end{array}
\right. ,
\end{equation}
which clearly has the symmetry
$\widetilde{H}^T(s)=\widetilde{H}(1-s)$. From this condition,
follows (\ref{symvp}) with $E=1$.

Let $w_{C\rightarrow C^{\prime }}^{eq}$ be some equilibrium
transition rates related to an equilibrium distribution $P^{eq}(C)$,
where detailed balance is satisfied. A sufficient condition for a
time-integrated current corresponding to transition rates
$w_{C\rightarrow C^{\prime }}$ to satisfy the GC symmetry
(\ref{ggc}), is that a quantity $E$ exists, such that these rates
take the form \cite{harris07}
\begin{equation}
w_{C\rightarrow C^{\prime }}=w_{C\rightarrow C^{\prime }}^{eq}\exp
(\frac{E}{2}\theta _{C\rightarrow C^{\prime }}).  \label{cs}
\end{equation}
Equivalently to (\ref{cs}), we have
\begin{equation}
w_{C\rightarrow C^{\prime }}=P_{eq}^{-1}(C)w_{C^{\prime }\rightarrow C}P_{eq}(C^{\prime })\exp (E\theta _{C\rightarrow C^{\prime }})  \label{cs'}
\end{equation}
Defining $P_{eq}$ as a diagonal matrix with elements $P_{eq}(C)$, it
then follows, from the last relation,
\begin{equation}
\tilde{H}^{T}(s)=P_{eq}^{-1}\tilde{H}(E-s)P_{eq},
 \label{c5PeqHPeq}
\end{equation}
which directly gives (\ref{symvp}). We remark that time-integrated
currents, such that an $E$ satisfying (\ref{cs}) can be found, are
proportional to the entropy (up to boundary terms) in the long time
limit (see \cite{lebowitz99,rakos08}).  More precisely,
\begin{equation}
\fl \Delta S_{m}=\sum_{k=1}^{N}\ln \frac{P^{eq}(C_{k})}{P^{eq}(C_{k-1})}+E\sum_{i=1}^{N}\theta _{C_{k-1}\rightarrow C_k}=\ln P^{eq}(C_{N})-\ln P^{eq}(C_{0})+EJ.
\end{equation}
Therefore, condition (\ref{cs}) is very restrictive in the sense
that very few currents satisfy it.

Nevertheless, if one considers the joint probability of more than
one time-integrated current a more general relation can be found
\cite{lebowitz99}. Let us take $k$ currents. Similar to (\ref{cs}),
a sufficient condition for a fluctuation relation including $k$
currents is that the rates take the form
\begin{equation}
w_{C\rightarrow C^{\prime }}=P_{eq}^{-1}(C)w_{C^{\prime }\rightarrow C}P_{eq}(C^{\prime })\exp \left( \sum_{i=1}^{k}E_{i}\theta _{C\rightarrow C^{\prime }}^{i}\right).
\label{cs2'}
\end{equation}
We define the large deviation function associated with the joint
probability distribution of the currents $J_{i}$ by
\begin{equation}
P(j_1,j_2,...,j_k) \approx  \exp (-\Delta TI(j_{1},j_{2},...,j_{k})).
\label{ratep}
\end{equation}
Analogously to (\ref{defmar}) we can define the deformed operator
\begin{equation}
\widetilde{H}_{CC'}(s_{1},s_{2},...s_{k})=\left\{
\begin{array}{c}
-w_{C^{\prime }\rightarrow C}\exp (-\sum_{i=1}^{k}s_{i}\theta _{C^{\prime
}\rightarrow C}^{i})\textrm{ if }C\neq C^{\prime } \\
\sum_{C^{\prime \prime }}w_{C\rightarrow C^{\prime \prime }}\textrm{ \ \ \ if }%
C=C^{\prime }
\end{array}
\right. ,
\end{equation}
and relation (\ref{cs2'}) gives
\begin{equation}
\tilde{H}^{T}(s_{1},s_{2},...s_{k})=P_{eq}^{-1}\tilde{H}(E_{1}-s_{1},E_{2}-s_{2},...,E_{k}-s_{k})P_{eq}.
\end{equation}
From the last relation we have a multidimensional version of
(\ref{symvp}), which reads
\begin{equation}
\widehat{I}(s_{1},s_{2},...s_{k})=\widehat{I}(E_{1}-s_{1},E_{2}-s_{2},...,E_{k}-s_{k}).
\end{equation}
In terms of the large deviation function we have the following
generalized GC symmetry \cite{lebowitz99},
\begin{equation}
I(j_1,j_2,...,j_k)-I(-j_1,-j_2,...,-j_k)= -\sum_{i=1}^kE_ij_i.
\label{FRseveral}
\end{equation}

\section{Fluctuations of the interface height}

Let us consider the RSOS model again. The physically meaningful
time-integrated current is the variation of height. Along the line
$p=1$ the height is just the entropy multiplying $\ln q$. Therefore,
for $p=1$ the probability distribution of the variation of height
displays the GC symmetry (\ref{ggc}) with $E=\ln q$. For $p\neq1$
the relation between height and entropy becomes more complicated,
and we would like to investigate the symmetry of the large deviation
function associated with the probability distribution of height for
this case. In other words, we would like to know if a symmetric
large deviation function (related to height) is a general property
of the RSOS model or if it holds only for specific microscopic rules
($p=1$). In the rest of this section we show that if a second
time-integrated current is considered, fluctuation relations for the
joint probability of two time-integrated currents that holds in the
whole phase diagram can be found. We then proceed to analyze the
symmetry of the Legendre transform of large deviation function
associated with height for small systems ($L=3,4$).

\subsection{Joint probability of two time-integrated currents}

We introduce the time-integrated currents $\Delta H_{q}$ and $\Delta
H_{qp}$ which are defined in the following way. The increment
$\theta_{C\rightarrow C^{\prime }}^{q}$ of $\Delta H_{q}$ is $-1$
for an evaporation taking place with rate $1$, $+1$ for the
deposition related to the reverse transition, and $0$ otherwise. In
a same way, the increment $\theta _{C\rightarrow C^{\prime }}^{qp}$
of $\Delta H_{qp}$ is $-1$ for an evaporation taking place with rate
$p$, $+1$ for the deposition related to the reverse transition and
$0$ otherwise. Note that, in the present case (\ref{cs2'}) is
verified on the form
\begin{equation}
w_{C\rightarrow C^{\prime }}=w_{C^{\prime }\rightarrow C}\exp \left( \left(\ln q\right) \theta _{C\rightarrow C^{\prime }}^{q}+\left( \ln \frac{q}{p}\right) \theta _{C\rightarrow C^{\prime }}^{qp}\right) .
\end{equation}
Then, relation (\ref{FRseveral}) becomes
\begin{equation}
I(h_q,h_{qp})-I(-h_q,-h_{qp})= -\ln (q)h_q-\ln  (q/p) h_{qp}
\label{hqphq}
\end{equation}
In order to include the height as a current we note that in terms of
$\Delta H_{q}$ and $\Delta H_{qp}$, the height variation (multiplied
by the system size $L$), $\Delta H$, is given by
\begin{equation}
\Delta H=  \Delta H_q +\Delta H_{qp}
\end{equation}
Hence, (\ref{hqphq}) implies in
\begin{eqnarray}
I(h,h_{qp})-I(-h,-h_{qp})= -\ln (q)h+\ln  (p) h_{qp}\nonumber\\
I(h,h_{q})-I(-h,-h_{q})=-\ln (q/p)h+\ln  (p) h_{q}.
\end{eqnarray}
The height appears explicitly in these relations and they are valid
in the entire phase diagram. Therefore, it is possible to find
fluctuation relations for the height if a second time-integrated
current is considered. Next, we consider the probability
distribution of the variation of height, without a second
time-integrated current.

\subsection{A symmetry different from the GC symmetry}

\begin{figure}[t]
\centering\includegraphics[width=120mm]{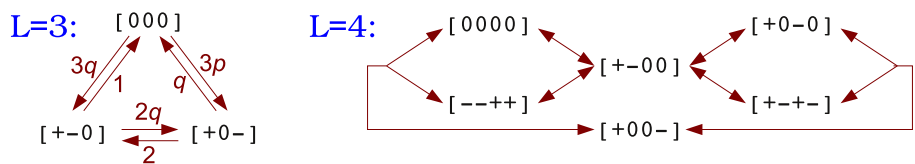}
\caption{Network of transitions in the charge representation of the RSOS model with $L=3$ (left) and $L=4$ sites (right).}
\label{fig:trnet}
\end{figure}

We consider the RSOS model for very small system sizes starting with
$L=3$, where the model has 7 states in the charge representation. In
order to visualize the network of transitions in this space, let us
denote by $[\sigma_1,\ldots,\sigma_L]$ all configurations which
differ only by translation, e.g. $[+-0]=\{+-0,\,0+-,\,-0+\}$. As
shown in the left panel of Fig.~\ref{fig:trnet}, the transitions for
$L=3$ form a simple cycle. Although each transition contributes
differently to $\Delta H$ and $\Delta S_{m}$ it is clear from
 the value of the rates (i.e. Fig.~\ref{fig:trnet}) that for
each completed cycle the resulting changes will have the fixed ratio
$\frac{\Delta S_{m}}{\Delta H}=\ln (q) -\frac13\ln (p)$. Therefore, in the long-time limit $\Delta
H$ will be distributed in the same way as $\Delta S_{m}$ and thus
trivially displays the GC symmetry. More precisely, it is possible
to find a similarity transformation of the form (\ref{c5PeqHPeq})
with $E=\ln (q) -\frac13\ln (p)$.

For $L=4$ the situation is different. As shown in the right panel of
Fig.~\ref{fig:trnet}, the transition network is a loop consisting of
two smaller cycles. If the system passes the whole loop it produces
both entropy and height, but if it circulates within the small
cycles it generates only entropy. This means that entropy and height
are no longer coupled and are expected to be distributed
differently.

To find out whether the distribution of $\Delta H$ is still
symmetric for $L=4$, let us follow the procedure described in the
previous section. In the charge basis the deformed evolution
operator $\tilde H(s)$ turns out to be a $19\times 19$ matrix of
which the lowest eigenvalue has to be found. The problem can be
simplified even further by defining the $6\times 6$ matrix
\begin{equation}
T_{\kappa\kappa'}(s) \;:=\; \sum_{c\in\kappa} \tilde H_{cc'}(s)\,,
\end{equation}
where $\kappa$ and $\kappa'$ denote the six sets of charge
configurations which differ only by translations and $C'$ is one of
the states represented by $\kappa'$. In the basis
$\{[0000],[--++],[+-00],[+-+-],[+0-0],[+00-]\}$, this matrix reads
\begin{equation}
\fl T_{\kappa\kappa'}(s) \;=\;
\left(\begin{array}{cccccc}
-4p-4q & 0 & e^s & 0 & 0 & qe^{-s} \\
0 & -1-q & pe^s & 0 & 0 & qe^{-s} \\
4qe^{-s} & qe^{-s} & -1-p-3q & 2e^s & 2e^s & 0 \\
0 & 0 & qe^{-s} & -2-2q & 0 & pe^s \\
0 & 0 & 2qe^{-s} & 0 & -2-2q & 2e^s \\
4pe^s & e^s & 0 & 2qe^{-s} & 2qe^{-s} & -2-p-2q
\end{array}\right)
\label{matrix}
\end{equation}
Since the eigenvector corresponding to the lowest eigenvalue of
$\tilde H(s)$ is translationally invariant, there will be a
corresponding eigenvector of $T(s)$ with the same eigenvalue. Hence
$\hat I(s)$ can be computed by determining the lowest eigenvalue of
the matrix given above.

\begin{figure}[t]
\centering\includegraphics[width=90mm]{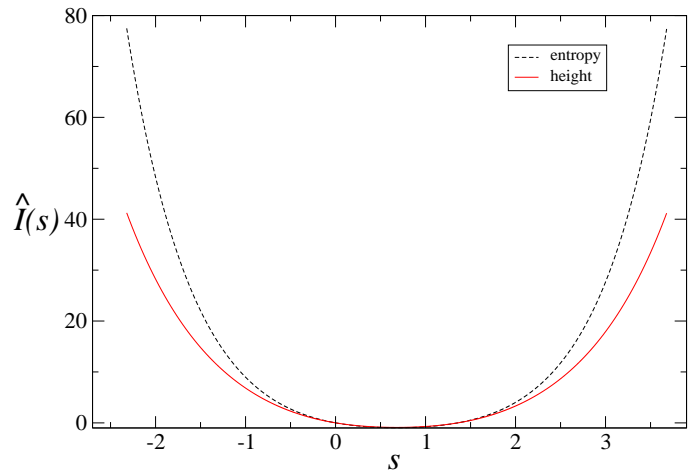}
\caption{Numerically determined Legendre transform $\hat{I}(s)$ of the large deviation function of the entropy (dashed line) rescaled by $s\to s/\frac{\ln 3q^4-\ln (2p+p^{2})}{4}$, and for the height (solid line) in the RSOS model with $p=0.1$ and $q=2$.
}
\label{fig:Iofs}
\end{figure}

Computing its determinant we find that $\hat I(s)$ vanishes for
$s=0$ and $E':=s=\frac14(\ln 3q^4-\ln(2p+p^2))$ and that the
characteristic polynomial and hence the spectrum is invariant under
the replacement $s\to E'-s$. Therefore, the RSOS model with $L=4$
sites has a symmetric large deviation function (associated with
height). Moreover, the function $\hat I(s)$ for the height current
-- even when appropriately rescaled -- differs from the
corresponding function for the entropy (see Fig.~\ref{fig:Iofs}),
showing that entropy and height are not proportional to each other
for $\Delta T\to \infty$. In Appendix C we present a different proof
for this symmetry, without using the matrix $\tilde H(s)$.

The GC symmetry has a clear physical interpretation. It is a result
of the fact that the exponential of the variation of the entropy for
a given stochastic path is the weight of the path divided by the
weight of the time-reversed path (see \cite{lebowitz99, harris07}).
Clearly, this is not the case of the height in the RSOS model,
therefore the symmetry we found for the $L=4$ case is different from
the GC symmetry. In Appendix $C$ we show that the symmetry is a
result of some subtle arrangement of the stochastic trajectories
that leads to a variation of height $\Delta H$. Hence, we have shown that currents different from the entropy may
also have a symmetric large deviation function.

Unfortunately, this observation does not extend to $L>4$, where a
small asymmetry in $\hat{I}(s)$ can be observed (we calculated
$\hat{I}(s)$ numerically, directly form the matrix $\tilde{H}(s)$,
for $L=5,6,7$). This may be caused by the increasingly complexity of
the network of transitions (see Appendix C).

\section{Conclusions}

In the present paper we studied the entropy production and
fluctuation relations for the particular solid-on-solid growth model
which belongs to the KPZ universality class. On a particular line of
the phase diagram, where the model can be solved exactly, we
calculated the entropy production, the velocity, and KPZ-parameter
$\lambda$. The results are compatible with the idea that $|\lambda|$
and $|v|$ can be used as a measure for the distance from
equilibrium.

Extending these exact results to the entire phase diagram within a
mean field approximation, we obtained analytical expressions for the
lines $v=0$ and $\lambda=0$ which are close to the true lines and
exhibit the same curvature. Furthermore, we calculated the entropy
production as a function of $\lambda$ along the line $v=0$. It was
found that $\dot{s}_m$ grows with $|\lambda|$, again in agreement
with the idea that the entropy production gives a measure of the
distance form criticality.

As we saw, along the line $p=1$ height is proportional to entropy
and therefore its probability distribution displays the GC symmetry.
We showed that if a second time-integrated current is considered a
symmetry in the joint probability distribution of two time
integrated currents valid in the whole phase diagram can be found.
Moreover, we explored the symmetry of the Legendre transform of the
large deviation function associated with the variation of height for
the $L=3,4$ cases. For $L=3$ the height turned out to be
asymptotically proportional to the entropy, hence it trivially
displays a GC symmetry. A different result emerged for $L=4$:
entropy and height are not proportional in the long time limit,
nevertheless the large deviation function related to height is still
symmetric. As we showed this  a new symmetry different from the GC
symmetry.

We would like to point out that the fluctuation relation for the
entropy production is very general, holding for any Markov process
with reversible transitions rates. Contrarily, it is valid for
very specific currents, that is, currents that become proportional
to the entropy in the long time limit. Here we found an example of a
current different from the entropy with a symmetric large deviation
function. It would be interesting to investigate what currents in
Markov processes, different from the entropy, have a symmetric large
deviation function and what may be the physical origin of the
symmetry.

{\noindent \textbf{Acknowledgements}}\newline The Deutsche
Forschungsgemeinschaft is gratefully acknowledged for partial
financial support (HI 744/3-1). We also thank the support of the
Israel Science Foundation (ISF) and the Minerva Foundation with
funding from the Federal German Ministry for Education and Research. 
RC acknowledges support from the Koshland Center for Basic Research.

\appendix
\section{Exact calculation of the stationary state of the RSOS model}

Here we provide more details on how the stationary state of the RSOS
model expressed in the charge basis can be computed exactly in a
certain subspace of the parameters by using the method introduced by
Neergaard and den Nijs~\cite{neergaard97}. Let us assume that the
probability to find the stationary system is proportional to
\begin{equation}
f_{M,E}=x^M y^E\,,
\end{equation}
where $x\leq 1$ and $y\leq 1$ are certain numbers depending
on the
rates and
\begin{equation}
M=\frac12\sum_j\sigma_j^2 \,, \qquad E=\sum_j j \sigma_j
\end{equation}
are the total number of the charges (divided by two) and their
electrostatic energy, respectively. Inserting this ansatz into the
master equation yields
\begin{eqnarray}
0 &=& n_{+-}[q_b f_{M-1,E+1}-p_b f_{M,E}]+ n_{-+}[p_a f_{M-1,E-1}-q_af_{M,E}]
\nonumber \\
&&+n_{00}[q_af_{M+1,E+1}+p_bf_{M+1,E-1}-(q_b+p_a)f_{M,E}]  \nonumber \\
&&+(n_{0-}+n_{+0})[q_cf_{M,E-1}-p_cf_{M,E}] \\
&&+(n_{-0}+n_{0+})[p_cf_{M,E+1}-q_cf_{M,E}]  \nonumber  \label{c5eqmaster}
\end{eqnarray}
where $n_{\sigma\sigma'}$ is the probability of finding a pair
$\sigma\sigma'$ of nearest neighbors sites in a configuration with
$M$ charges and electrostatic energy $E$. Using the relation
\begin{equation}
n_\sigma\;=\;\sum_{\sigma'}n_{\sigma\sigma'} \;=\;
\left\{\begin{array}{ll}
M/L & \mbox{ if } \sigma=\pm 1\\
1-2M/L &  \mbox{ if } \sigma=0
\end{array}
\right.
\end{equation}
one obtains the constraint
\begin{equation}
n_{0-}+ n_{+0}+2n_{+-} \;=\; n_{-0}+ n_{0+}+ 2n_{-+} \,.
\end{equation}
meaning that there are four independent variables in the master
equation (\ref{c5eqmaster}), namely $n_{00}$, $n_{+-}$, $n_{-+}$ and
$(n_{0-}+n_{+0}+2n_{+-})$, and that the corresponding coefficients
have to vanish. The coefficient of $(n_{0-}+ n_{+0}+2n_{+-})$
vanishes if
\begin{equation}
(q_c-p_cy^{-1})(1-y)= 0.
\end{equation}
This equation has two solutions. The first one, $y=q_c/p_c$,
combined with the other constraints, reduces to the condition of
detailed balance (see~\cite{neergaard97}). The second solution
$y=1$, on which we will focus in the following, is much more
interesting because it extends to the nonequilibrium regime of the
model. In this case the configurational probability distribution
becomes independent of the electrostatic energy $E$.

For this solution the coefficient of $n_{00}$ gives
\begin{equation}
x= \frac{q_b+p_a}{q_a+p_b}.  \label{c5eqx}
\end{equation}
The two remaining coefficients of $n_{+-}$ and $n_{-+}$ lead to the
same relation, namely
\begin{equation}
q_c-p_c= \frac{q_bq_a-p_ap_b}{2(q_b+p_a)}.  \label{c5eqconstraint}
\end{equation}
Therefore, in the region of the phase space where the constraint
(\ref{c5eqconstraint}) is satisfied the probability of a
configuration with $M$ charges in system of size $L$ in the
stationary state is exactly given by
\begin{equation}
P(M)= Z_{M,L}^{-1}x^{M}  \label{c5eqprobN}
\end{equation}
where
\begin{equation}
Z_{M,L}= \sum_{M=0}^{L/2}\frac{L!}{M!M!(L-2M)!}x^{M}  \label{c5eqpart}
\end{equation}
is the partition function and $x$ is given by (\ref{c5eqx}).
In finite systems neighboring sites are still correlated because of charge conservation. For example, in a system with 2 sites the only possible configurations are 00, +-, and -+ and thus their probability distribution does not factorize. However, in the limit $L\to\infty$ this conservation law will be less and less important, meaning that the state factorizes asymptotically:
\begin{equation}
\label{eq:factorize}
\langle\sigma\sigma'\rangle=\langle\sigma\rangle
\langle\sigma'\rangle+o(1/L),  \label{c5factor}
\end{equation}
Moreover, in the large $L$ limit the sum in (\ref{c5eqpart})
is dominated by its maximum and the
density of charges $\rho= 2M/L$ is given by
\begin{equation}
\rho= \frac{2}{2+x^{-1/2}}.  \label{c5eqrho}
\end{equation}

\section{Mean field calculation of $\lambda$ and $\dot s_m$}

To compute the value of the coefficient $\lambda$ in the KPZ
equation one has to consider a tilted interface in the stationary
state and to investigate how the interface velocity $v$ varies with
the slope $m$. According to equation (\ref{c5eqlambda2}), $\lambda$
is then given by
\begin{equation}
\label{eq:lambda}
\lambda \;=\; \frac{1}{m}\,\frac{d}{dm} \, v(m)
\end{equation}
where $v(m)$ is given by Eq.~(\ref{eq:velocity}).

To compute $v(m)$ we assume that $\sigma_i\sigma_{i+1}$ factorizes.
Of course this is only correct for $m=0$ in the parameter subspace
constrained by (\ref{c5eqconstraint}). However, it is reasonable to
expect that the stationary state still factorizes if $m$ is
sufficiently small. With this assumption we have
\begin{equation}
\ex+ =\frac{\rho(m)+m}2 \,, \quad
\ex- =\frac{\rho(m)-m}2 \,, \quad
\ex 0=1-\rho(m)
\end{equation}
with some unkown function $\rho(m)$ for the charge density, so that
the interface velocity~(\ref{eq:velocity}) is given by
\begin{eqnarray}
 v(m) &=& (q_b-p_a)\bigl(1-\rho(m)\bigr)^2+\frac{1}{4}(q_a-p_b)\bigl(\nonumber
\rho(m)^2-m^2\bigr) \\ && +\, (q_c-p_c)\bigl(1-\rho(m)\bigr)\rho(m);  \label{c5eqvelo}
\end{eqnarray}
To compute $\lambda$ we need to know the derivative of $\rho(m)$.
Since
\begin{equation}
\frac{d}{dt}\rho(t)=
(q_b+p_a)\langle00\rangle-q_a\langle-+\rangle-p_b\langle+-\rangle
\label{c5defrho}
\end{equation}
vanishes in the stationary state, we have
\begin{equation}
(q_b+p_a)(1-\rho(m))^2-\frac{1}{4}(q_a+p_b)(\rho(m)^2-m^2)\;=\;0\,.
\label{eqB5}
\end{equation}
Differentiating with respect to $m$ we obtain the expression
\begin{equation}
m^{-1}\frac{d}{dm}\rho(m) \;=\; [\rho+4\frac{(q_b+p_a)}{q_a+p_b}(1-\rho)]^{-1}.
\label{c5eqden}
\end{equation}
which tends to $\frac1{2x}$ in the limit $m\to 0$. Inserting this
result into (\ref{eq:lambda}) yields
\begin{equation}
\fl\lambda \;=\; \frac{x^{-1/2}}{4}[(q_a-p_b)\rho-4(q_b-p_a)(1-%
\rho)+2(q_c-p_c)(1-2\rho)]-\frac{(q_a-p_b)}{2}.  \label{c5eqlambda}
\end{equation}
One can obtain the same expression for $\lambda$ by differentiating
the deterministic part of the KPZ equation for the RSOS
model~\cite{neergaard97}.Similarly, we can compute the entropy
production rate~(\ref{eq:entropypair}):
\begin{equation}
\fl \dot{s}_m = (q_a\ln\frac{q_a}{p_a}-p_b\ln\frac{q_b}{p_b})\frac{\rho^2}{4}
+(q_b\ln\frac{q_b}{p_b}-p_a\ln\frac{q_a}{p_a})(1-\rho)^2
+ (q_c-p_c)\ln\frac{q_c}{p_c}\rho(1-\rho).  \label{c5eqentprod}
\end{equation}
Thus we have obtained the quantities $v$, $\dot{s}_m$ and $\lambda$ as
functions of the density of charges $\rho$ in the stationary state, assuming
factorization of $\langle\sigma\sigma'\rangle$. These results are exact
in the parameter subspace constrained by (\ref{c5eqconstraint}) in the limit $L\to \infty$,
where the density of charges is given by (\ref{c5eqrho}).

\section{The special symmetry  for $L=4$}

In this Appendix we show that the large deviation function
associated with the height variation $I(h)=-\lim_{\Delta T \to
\infty} \frac{1}{\Delta T}\ln P(h)$, where $h=\Delta H/ \Delta T$,
has the symmetry
\begin{equation}
I(h)-I(-h)= -h\frac14\ln\frac{3q^4}{2p+p^2},
\label{first}
\end{equation}
for the RSOS model with $L=4$.

We start by considering the embedding Markov chain (discrete time) of the Markovian process. The
transition matrix reads
\begin{equation}
T_{C\rightarrow C^{\prime }}=\frac{w_{C\rightarrow C^{\prime }}}{\lambda _{C}%
}
\end{equation}
where $\lambda _{C}=\sum_{C^{\prime }\neq C}w_{C\rightarrow
C^{\prime }}$. Within the basis used in (\ref{matrix}), the
transition matrix takes the explicit form
\begin{equation}
T=\left(
\begin{array}{cccccc}
0 & 0 & \frac{1}{3q+p+1} & 0 & 0 & \frac{q}{2q+p+2} \\
0 & 0 & \frac{p}{3q+p+1} & 0 & 0 & \frac{q}{2q+p+2} \\
\frac{q}{q+p} & \frac{q}{q+1} & 0 & \frac{1}{q+1} & \frac{1}{q+1} & 0 \\
0 & 0 & \frac{q}{3q+p+1} & 0 & 0 & \frac{p}{2q+p+2} \\
0 & 0 & \frac{2q}{3q+p+1} & 0 & 0 & \frac{2}{2q+p+2} \\
\frac{p}{q+p} & \frac{1}{q+1} & 0 & \frac{q}{q+1} & \frac{q}{q+1} & 0
\end{array}
\right)  \label{T}
\end{equation}
In this discrete time case we are interested in the probability
of having a variation of height $\Delta H$ with a fixed number of
jumps $N$, which we denote by $P(\left. \Delta H\right| N)$. Our aim
is to obtain all possible paths that lead to a variation of height
$\Delta H$ within $N$ jumps. Let us assume that $\Delta H$ and $N$
are large, such that only periodic sequence of jumps contribute,
while any other sequence of jumps will just generate a boundary term
that is irrelevant for large $\Delta H$ and $N$. We also consider an
initial state $[+-00]$, which does not impose any restriction since
we assume $\Delta H$ and $N$ large.

Among the possible periodic sequence of jumps, there are four cycles
that increase the height. They are:
\begin{enumerate}
\item  {\ $[+-00]\rightarrow \lbrack +0-0]\rightarrow \lbrack
+00-]\rightarrow \lbrack 0000]\rightarrow \lbrack +-00]$ }

\item  {\ $[+-00]\rightarrow \lbrack +-+-]\rightarrow \lbrack
+00-]\rightarrow \lbrack 0000]\rightarrow \lbrack +-00]$ }

\item  {\ $[+-00]\rightarrow \lbrack +0-0]\rightarrow \lbrack
+00-]\rightarrow \lbrack --++]\rightarrow \lbrack +-00]$ }

\item  {\ $[+-00]\rightarrow \lbrack +-+-]\rightarrow \lbrack
+00-]\rightarrow \lbrack --++]\rightarrow \lbrack +-00]$ }
\end{enumerate}

These cycles take $4$ jumps each and increase height by $4$.
Defining $A=(3q+p+1)(q+1)(q+p)(2+2q+p)$ and
$B=(3q+p+1)(q+1)^{2}(2+2q+p)$, the probability of cycle $1$ is
$M_{1}=2q^{4}/A,$ of cycle $2$ is $M_{2}=q^{4}/A,$ of cycle 3 is
$M_{3}=2q^{4}/B$ and of cycle 4 is $M_{4}=q^{4}/B$. The reversed
cycles decrease height by 4 and they have probabilities
$\widetilde{M_{1}}=p^{2}/A$ for cycles $1$,
$\widetilde{M_{2}}=2p/A${\ \ for cycles $2,$
}$\widetilde{M_{3}}${\ $=p^{2}/B$ for cycles }$3$ and
$\widetilde{M_{4}}=2p/B$\ $\ $for cycle 4. We denote the number
of these cycles in some path ${C}= C_0\to C_1\to ... \to C_N$ by
$m_{i}$ and the number of reversed cycles by $\tilde{m}_{i}$, with
$i=1,2,3,4$.

Besides these four cycles (and their four reversed ones) there are
many other periodic sequence of jumps, however they do not change
height. For example, it is possible  to jump from $[+0-0]$ to
$[+00-]$ and then jump back. This sequence takes two jumps and it
does not change height. Another example, is the cycle
$[+-00]\to[+0-0]\to[+00-]\to[+-+-]\to[+-00]$, which takes four jumps
and also does not change height. We denote by $\Delta $  the set of
independent periodic sequences of jumps that do not change height.
Each element of the set,  $i\in \Delta$, is characterized by its
probability $P_{i}$ and the number of jumps it takes $\eta _{i}$.
The random number $n_{i}$ gives the number of time the sequence
$i\in \Delta $ appear in the path $\{C\}$.

The probability of the variation of height $\Delta H$ with a fixed
number of jumps $N$ can be written in the form
\begin{equation}
P(\left. \Delta H\right| N)=\textrm{$\sum_{\{C\}}'$}W[\{C\}],
\label{sum}
\end{equation}
where $W[\{C\}]$ is the probability of the path $\{C\}$ and the sum $\sum_{\{C\}}'$
is over all paths with $N$ jumps and variation of height $\Delta H$.
By commutativity of the measure of the stochastic path, one can
order the terms contributing to the probability of a path such that
a given trajectory looks like a sequence of elements of $\Delta$,
followed by a sequence of cycle $1$, $2,$ $3$ or $4$ (or their
reversed). If we have
$\tilde{m_1}+\tilde{m_2}+\tilde{m_3}+\tilde{m_4}=j$, then
$m_1+m_2+m_3+m_4=\Delta H/4+j$ and ${\sum_{i\in \Delta }} n_{i}\eta
_{i}{=N-\Delta H-8j}$. Therefore, the sum (\ref{sum}) becomes
\begin{eqnarray}
\sum_{j=0}^{(N-\Delta H)/8}\left( \sum_{\left\{ \eta _{i}\right\} }\delta \left( {
\sum_{i\in \Delta }}n_{i}\eta _{i}-{N+\Delta H+8j}\right) \left( \prod_{i\in
\Delta }{P_{i}^{n_{i}}}\right) \right)\times\nonumber\\
 \left( M_{1}+M_{2}+M_{3}+M_{4}\right)
^{{\Delta H/4+j}}\left( \widetilde{M_{1}}+\widetilde{M_{2}}+\widetilde{M_{3}}+
\widetilde{M_{4}}\right) ^{j}.
\label{expPH}
\end{eqnarray}
where the term $(M_1+ M_2+ M_3+ M_4)^{\Delta H/4+j}$ comes form the
sum over all possible $m_1, m_2, m_3, m_4$ subjected to the
constraint   $m_1+m_2+m_3+m_4=\Delta H/4+j$ and, similarly, the term
$(\widetilde{M_1}+ \widetilde{M_2}+ \widetilde{M_3}+
\widetilde{M_4})^{j}$ comes form the sum over all possible
$\widetilde{m_1}, \widetilde{m_2}, \widetilde{m_3}, \widetilde{m_4}$
subjected to the constraint
$\widetilde{m_1}+\widetilde{m_2}+\widetilde{m_3}+\widetilde{m_4}=j$.
Analogously, for $P(\left. -\Delta H\right| N)$ we obtain
\begin{eqnarray}
\sum_{j=0}^{(N-\Delta H)/8}\left( \sum_{\left\{ \eta _{i}\right\} }\delta \left( {
\sum_{i\in \Delta }}n_{i}\eta _{i}-{N+\Delta H+8j}\right) \left( \prod_{i\in
\Delta }{P_{i}^{n_{i}}}\right) \right) \times\nonumber\\
\left( M_{1}+M_{2}+M_{3}+M_{4}\right)
^{{j}}\left( \widetilde{M_{1}}+\widetilde{M_{2}}+\widetilde{M_{3}}+
\widetilde{M_{4}}\right) ^{{\Delta H/4+j}}.
\label{expPH2}
\end{eqnarray}
Therefore, the probability of the variation of height with a fixed
number of jumps has the symmetry
\begin{equation}
\fl \frac{ P(\left. \Delta H \right| N)}{
P(\left. -\Delta H\right| N)}=\left( \frac{%
M_{1}+M_{2}+M_{3}+M_{4}}{\widetilde{M_{1}}+\widetilde{M_{2}}+\widetilde{M_{3}%
}+\widetilde{M_{4}}}\right) ^{{\Delta H/4}}.
\end{equation}
Note that right hand side of the last equation does not depend on
the number of jumps $N$. By applying Bayes formula, we have
\begin{equation}
\frac{P(h)}{P(-h)}= \frac{\sum_{N=\left\lceil Th\right\rceil} P(\left. \Delta H \right| N)P(N)}
{\sum_{N=\left\lceil Th\right\rceil } P(\left.-\Delta  H \right| N)P(N)}=
\left( \frac{M_{1}+M_{2}+M_{3}+M_{4}}{\widetilde{M_{1}}+\widetilde{M_{2}}+\widetilde{M_{3}}+\widetilde{M_{4}}}\right) ^{\frac{T}{4}h},
\label{last}
\end{equation}
where $h= \Delta H/\Delta T$, $P(h)$ is the probability in the
continuous time process with fixed $\Delta T$ and $P(N)$ is the
probability of having $N$ jumps.

Finally, since
\begin{equation}
\fl M_{1}+M_{2}+M_{3}+M_{4}=3q^{4}\left( \frac{1}{A}+\frac{1}{B}\right)\textrm{ and } \widetilde{M_1}+\widetilde{M_2}+\widetilde{M_3}+\widetilde{M_4}=( p^2+2p) ( \frac{1}{A}+\frac{1}{B}),
\end{equation}
from (\ref{last}) we obtain the symmetry (\ref{first}).

An important point in the calculation we performed here is that the
chosen initial state $[+-00]$ has the property that any cycle that
increases height  has to go through it (we could also choose
$[+00-]$). Without this property it would not be possible to write
down equations (\ref{expPH}) and (\ref{expPH2}), from which we
derive the symmetry. As we mentioned, the symmetry is lost for $L=5$
where the large deviation function displays a small asymmetry. By
drawing the network of states for the $L=5$ case, which has $10$
states, we observed that there is no state such that a cycle that
increases height has to go trough it.

\end{document}